\documentstyle[12pt,psfig]{article} 
\makeatletter \def\@cite#1#2{{[{#1}]\if@tempswa\typeout {IJCGA
warning: optional citation argument ignored: `#2'} \fi}}

\newcount\@tempcntc
\def\@citex[#1]#2{\if@filesw\immediate\write\@auxout{\string\citation{#2}}\fi
  \@tempcnta\z@\@tempcntb\m@ne\def\@citea{}\@cite{\@for\@citeb:=#2\do
    {\@ifundefined
       {b@\@citeb}{\@citeo\@tempcntb\m@ne\@citea\def\@citea{,}{\bf ?}\@warning
       {Citation `\@citeb' on page \thepage \space undefined}}%
    {\setbox\z@\hbox{\global\@tempcntc0\csname b@\@citeb\endcsname\relax}%
     \ifnum\@tempcntc=\z@ \@citeo\@tempcntb\m@ne
       \@citea\def\@citea{,}\hbox{\csname b@\@citeb\endcsname}%
     \else
      \advance\@tempcntb\@ne
      \ifnum\@tempcntb=\@tempcntc
      \else\advance\@tempcntb\m@ne\@citeo
      \@tempcnta\@tempcntc\@tempcntb\@tempcntc\fi\fi}}\@citeo}{#1}}
\def\@citeo{\ifnum\@tempcnta>\@tempcntb\else\@citea\def\@citea{,}%
  \ifnum\@tempcnta=\@tempcntb\the\@tempcnta\else
   {\advance\@tempcnta\@ne\ifnum\@tempcnta=\@tempcntb \else 
\def\@citea{--}\fi
    \advance\@tempcnta\m@ne\the\@tempcnta\@citea\the\@tempcntb}\fi\fi}
\makeatother

\def\boxit#1{\leavevmode\thinspace\hbox{\vrule\vtop{\vbox{\hrule%
        \vskip3pt\kern1pt\hbox{\vphantom{\bf/}\thinspace\thinspace%
        {\bf#1}\thinspace\thinspace}}\kern1pt\vskip3pt\hrule}\vrule}%
        \thinspace}
\def\Boxit#1{\noindent\vbox{\hrule\hbox{\vrule\kern3pt\vbox{
\advance\hsize-7pt\vskip-\parskip\kern3pt\bf#1 \hbox{\vrule height0pt
depth\dp\strutbox width0pt} \kern3pt}\kern3pt\vrule}\hrule}}

\newcommand{\Hh}{\lower1.2ex\hbox{$\stackrel{\textstyle
H}{\footnotesize\sim}$}}
\newcommand{\Hho}{\lower1.2ex\hbox{$\stackrel{\textstyle
H_1}{\footnotesize\sim}$}}
\newcommand{\Hhw}{\lower1.2ex\hbox{$\stackrel{\textstyle
H_2}{\footnotesize\sim}$}}
\newcommand{\h}{\lower1.2ex\hbox{$\stackrel{\textstyle
h}{\footnotesize\sim}$}}

\newcommand{\gsim}{\lower.7ex\hbox{$\;\stackrel{\textstyle>}{\sim}\;$}}
\newcommand{\lsim}{\lower.7ex\hbox{$\;\stackrel{\textstyle<}{\sim}\;$}}
\newcommand{\be}{\begin{equation}} \newcommand{\ee}{\end{equation}}
\newcommand{\beq}{\begin{equation}} \newcommand{\eeq}{\end{equation}}
\newcommand{\bea}{\begin{eqnarray}} \newcommand{\eea}{\end{eqnarray}}
\newcommand{\SUSY}{\makebox[1.3cm][l]{$\line(4,1){35}$\hspace{-1.15cm}{SUSY}}}
\newcommand{\tinySUSY}{\makebox[0.85cm][l]{$\line(4,1){19}$\hspace{-0.77cm}
{\tiny{SUSY}}}} \def\bma#1{\mbox{\boldmath{$#1$}}}
\def\simlt{\stackrel{<}{{}_\sim}} \def\simgt{\stackrel{>}{{}_\sim}}

\def\baselinestretch{1}
\begin{document}
\catcode`@=11 \newtoks\@stequation
\def\subequations{\refstepcounter{equation}%
\edef\@savedequation{\the\c@equation}%
\@stequation=\expandafter{\theequation}
\edef\@savedtheequation{\the\@stequation}
\edef\oldtheequation{\theequation}
\def\theequation{\oldtheequation\alph{equation}}}
\def\endsubequations{\setcounter{equation}{\@savedequation}%
\@stequation=\expandafter{\@savedtheequation}%
\edef\theequation{\the\@stequation}\global\@ignoretrue

\noindent} \catcode`@=12
\begin{titlepage}

\title{{\bf Implications for New Physics\\ from Fine-Tuning
Arguments:\\ I. Application to SUSY and Seesaw Cases}} \vskip3in \author{   
{\bf
J.A. Casas},
{\bf
J.R. Espinosa}
and  {\bf
I. Hidalgo\footnote{\baselineskip=16pt E-mail addresses: {\tt
alberto.casas@uam.es, jose.espinosa@cern.ch, irene.hidalgo@uam.es}}}
\hspace{3cm}\\
{\small IFT-UAM/CSIC, 28049 Madrid, Spain}.
}  \date{}  \maketitle  \def\baselinestretch{1.15}
\begin{abstract}
\noindent
We revisit the standard argument to estimate the scale of new physics (NP)
beyond the SM, based on the sensitivity of the Higgs mass to quadratic
divergences. Although this argument is arguably naive, the corresponding
estimate, $\Lambda_{\rm SM}\simlt 2-3$ TeV, works reasonably well in most
cases and should be considered a conservative bound, as it ignores
other contributions to the Higgs mass which are potentially large. 
Besides, the
possibility of an accidental Veltman-like cancellation does not raise
significantly $\Lambda_{\rm SM}$. One can obtain more precise implications 
from
fine-tuning arguments in specific examples of NP. Here we consider SUSY
and right-handed (seesaw)  neutrinos.  SUSY is a typical example for which
the previous general estimate is indeed conservative: the MSSM is
fine-tuned a few \%, even for soft masses of a few hundred
GeV.  In contrast, other SUSY scenarios, in particular those with 
low-scale SUSY
breaking, can easily saturate the general bound on $\Lambda_{\rm SM}$. The
seesaw mechanism requires large fine-tuning if $M_R\simgt 10^7$ GeV,
unless there is {\em additional} NP (SUSY being a favourite option).
\end{abstract}

\thispagestyle{empty}
\vspace*{1.cm} \leftline{October 2004} \leftline{}

\vskip-19cm \rightline{IFT-UAM/CSIC-04-57} 
\rightline{hep-ph/0410298} \vskip3in

\end{titlepage}
\setcounter{footnote}{0} \setcounter{page}{1}
\newpage
\baselineskip=20pt

\noindent

\section{Introduction}

It is commonly assumed that the Big Hierarchy Problem of the Standard
Model (SM) indicates the existence of New Physics (NP) beyond the SM at
a scale  $\Lambda_{\rm SM}\simlt$ few TeV. The argument is quite
simple and well known. In the SM (treated as an effective theory valid
below $\Lambda_{\rm SM}$) the mass parameter in the Higgs potential, $m^2$, 
receives important
quadratically-divergent contributions. At one-loop \cite{veltman},
\be
\label{quadrdiv0}
\delta_{\rm q} m^2 = {3\over 64\pi^2}(3g^2 + g'^2 + 8\lambda -
8\lambda_t^2)\Lambda_{\rm SM}^2\ , 
\ee
where $g, g', \lambda$ and $\lambda_t$ are the $SU(2)\times U(1)_Y$
gauge couplings, the quartic Higgs coupling and the top Yukawa coupling 
respectively. In terms of masses
\be
\label{quadrdiv}
\delta_{\rm q} m^2 = {3\over 16\pi^2 v^2}(2m_W^2 + m_Z^2 + m_h^2 -
4 m_t^2)\Lambda_{\rm SM}^2\ , 
\ee
where $m_W^2=g^2v^2/4$, $m_Z^2=(g^2+g'^2)v^2/4$, $m_h^2=2\lambda v^2$ and 
$m_t^2=\lambda_t^2v^2/2$, with $v=246$ GeV. Upon minimization of the 
potential, this translates into a dangerous contribution to the Higgs 
vacuum expectation value (VEV) which tends to destabilize the electroweak 
scale.
The requirement of no fine-tuning between the 
above contribution
and the tree-level value of $m^2$ sets an upper bound on
$\Lambda_{\rm SM}$. E.g. for $m_h = 115-200$ GeV
\be
\label{quadrft}
\left|{\delta_{\rm q} m^2 \over m^2}\right|=
\left|{\delta v^2 \over v^2}\right|\leq 10\ \Rightarrow \ 
\Lambda_{\rm SM}  \simlt 2-3\ {\rm TeV} \ ,
\ee
where we have used $m_h^2=2 m^2$ [valid in the approximation of 
disregarding terms beyond ${\cal O}(h^4)$ in the Higgs potential].
This upper bound on $\Lambda_{\rm SM}$ is in a certain tension with
bounds on the suppression scale of higher order operators derived from 
fits to precision electroweak data \cite{ewfits}, which require
$\Lambda\simgt$ 10 TeV or even higher for some particularly dangerous
operators (e.g. if they are flavour non-conserving).
This is known as the ``Little Hierarchy'' problem, and implies that  the 
NP at $\Lambda_{\rm SM}$
should be ``clever'' enough to be consistent with these constraints,
giving some hints about its properties.
E.g. non-strongly-interacting, flavour-blind NP is favoured.

\begin{figure}[t]
\vspace{1.cm} \centerline{
\psfig{figure=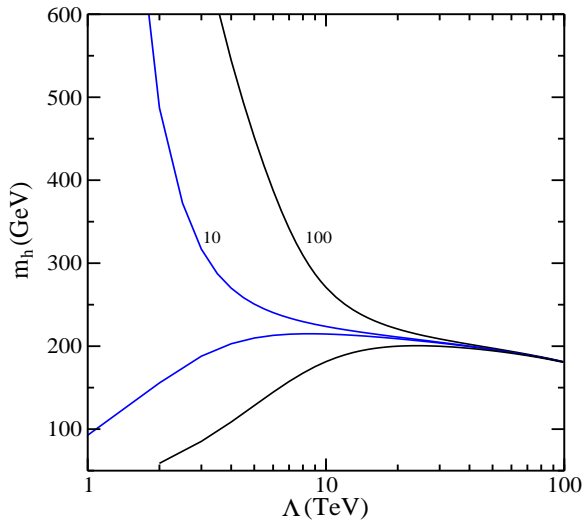,angle=-90,height=6cm,width=9cm,bbllx=4.cm,%
bblly=0.cm,bburx=20.cm,bbury=23.cm}
}
\caption{\footnotesize
Fine-tuning contours corresponding to $\Delta_{\Lambda}=10$ and 100, i.e. 
10\% and 1\% fine-tuning, respectively.}
\label{fig:velt-a}
\end{figure}

The previous ``Big Hierarchy'' argument leads to an optimistic prospect, 
as it sets the scale of NP on the reach of LHC. However, 
as stressed in ref.~\cite{KM}, $\Lambda_{\rm SM}$ could be much larger if 
the Higgs mass 
lies (presumably by
accident) close to the value that cancels $\delta_{\rm q}
m^2$ in (\ref{quadrdiv}). At one-loop this is the 
famous Veltman's condition
\cite{veltman}
\be
\label{veltman}
2m_W^2 + m_Z^2 + m_h^2 - 4 m_t^2 \simeq 0\ , \ee
which is fulfilled for $m_h\simeq 307$ GeV \footnote{Extended models 
constructed to lower this value  have been considered in 
ref.~\cite{Calmet}.}. 
At higher order the condition  becomes cut-off dependent \cite{EJ}
\be
\label{veltmanEJ}
\sum_{n\geq 0}c_n(\lambda_i)\log^n{\Lambda \over m_h}\simeq 0\ , \ee
where $\lambda_i$ are the couplings of the SM and $c_0(\lambda_i) =(3g^2 +
g'^2 + 8\lambda - 8\lambda_t^2)$, the combination that appears in the
one-loop Veltman correction. Eq.~(\ref{veltmanEJ}) is nothing but
$c_0(\lambda_i)$ evaluated at the cut-off scale $\Lambda$ and therefore
contains the leading-log corrections to all orders. However, it does not
include subleading (finite and next-to-leading-log) contributions which
are not important for moderate values of $\Lambda$ \cite{KM}. The results
of ref.~\cite{KM}, obtained by working at two-loop leading-log, are
summarized in the fig.~2 of that paper, which is similar to our
fig.~\ref{fig:velt-a}, to be discussed in the next section. There, the
lines of 10\% and 1\% fine-tuning show a throat (sometimes called
``Veltman's throat''), which corresponds to eq.~(\ref{veltmanEJ}). More
precisely, the plot indicates that if $m_h\sim 210-225$ GeV, $\Lambda_{\rm
SM}$ could be $\simgt 10$ TeV with no fine-tuning price and no Little
Hierarchy problem. Incidentally, this solution would be far more economic
than Little Higgs models \cite{LH}. (Likewise, a Little Higgs model with
$m_h$ in the above range would lack genuine motivation.) On the negative
side, this means that if $m_h$ is in this range, NP might escape detection
at LHC.

In this paper we re-examine the use of the Big Hierarchy Problem of the SM
to extract information about the size of $\Lambda_{\rm SM}$, illustrating
our general results with physically relevant examples.

In section 2 we consider the possibility of living near a Veltman's
condition. In contrast with previous discussions in the literature, we
show that this potential circumstance does not reduce the amount of
fine-tuning (once it is evaluated in a complete way), so that the upper 
bound on $\Lambda_{\rm SM}$ is not affected much at present.

In section 3, we examine the limitations of the Big Hierarchy argument to
estimate the scale of new physics. We argue that the reasoning, as usually
presented and summarized above, is too naive. However, quantitatively, the
corresponding estimate for the upper bound on $\Lambda_{\rm SM}$ turns out
to work reasonably well in most cases, and, indeed, it should be
considered a conservative bound, as it ignores potentially large
contributions to the Higgs mass parameter.

In section 4, we analyze two physically relevant examples of NP:  
right-handed (seesaw) neutrinos and supersymmetry (SUSY).  We discuss how,
in the context of the SM, the seesaw mechanism produces a very important
fine-tuning problem which claims for the existence of {\em additional} NP.
On the other hand SUSY illustrates the fact that the ``naive'' bound is
indeed conservative. E.g. the MSSM is quite fine-tuned even for soft
masses of a few hundred GeV. We discuss how other SUSY scenarios, in
particular those with low-scale SUSY breaking, can easily evade the
problematic aspects of the MSSM, essentially saturating the general bound.

Another example of new physics, namely Little Higgs models, deserve a 
separate analysis and will be the subject of a companion paper.

Finally, we summarize and  present our conclusions in section 5.

\section{Remarks on the shape of Veltman's throat}

We start our analysis by re-examining the
argument that led to the expectation of a fine-tuning throat, as
presented in the previous section. 
To quantify the tuning we follow Barbieri and Giudice \cite{BG}:
we write the Higgs VEV as $v^2=v^2(p_1, p_2, \cdots)$, where $p_i$ are
initial parameters of the model under study, and 
measure the amount of fine tuning associated to $p_i$ by
$\Delta_{p_i}$, defined as
\be
\label{ftBG}
{\delta M_Z^2\over M_Z^2}= {\delta v^2\over v^2} = \Delta_{p_i}{\delta
p_i\over p_i}\ ,  
\ee
where $\delta M_Z^2$ (or $\delta v^2$) is the change induced in
$M_Z^2$ (or $v^2$) by a change $\delta p_i$ in $p_i$. Roughly
speaking $\Delta^{-1}_{p_i}$ measures the probability of a 
cancellation among terms of a
given size to obtain a result which is $\Delta_{p_i}$ times smaller. 

The discussion of the previous section concerns the dependence of
$v^2$ on the scale $\Lambda_{\rm SM}$. Indeed, plotting  the
lines of $\Delta_\Lambda =$ 10, 100 in the ($m_h, \Lambda$) plane, as
shown in fig.~\ref{fig:velt-a}, we obtain curves similar to the 10\% and 
1\% curves in fig.~2 of ref.~\cite{KM}\footnote{We compute the whole 
leading log contribution to $\delta_{\rm q} m^2$, using 
eq.~(\ref{veltmanEJ}).}, as discussed in the Introduction, 
reproducing the throat already mentioned.

Our first remark has to do with the fact that, besides $\Lambda$, there
are other relevant parameters which are not yet measured with extreme
precision. The most notable case is the top mass, which, according to the
latest experimental data \cite{top}, is 
\be
\label{topmass}
M_t = 178 \pm 4.3 \ {\rm GeV}\ . \ee
%
\begin{figure}[t]
\vspace{1.cm} \centerline{
\psfig{figure=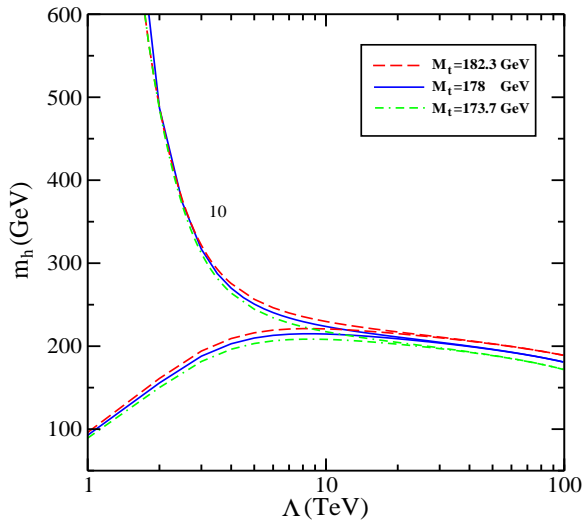,angle=-90,height=6cm,width=9cm,bbllx=4.cm,%
bblly=0.cm,bburx=20.cm,bbury=23.cm}
}
\caption{\footnotesize
Same as fig.~\ref{fig:velt-a} for $\Delta_{\Lambda}=10$ and three 
different values 
of the top mass.
}
\label{fig:velt-b}
\end{figure}
%
%
\noindent 
Although this uncertainty is remarkably small, it should not be ignored
for fine-tuning issues. Figure~\ref{fig:velt-b} shows three curves with 
$\Delta_\Lambda =$ 10, 
corresponding to $M_t =$ 173.7, 178 and 182.3 GeV. 
The value of $\Delta_\Lambda$ should be averaged over all the allowed 
ranges of variation of the unknown parameters\footnote{This can be 
understood by noting that, for a fixed value of $M_t$, $\Delta_\Lambda$ 
has the statistical meaning of $\Delta_\Lambda\sim 
\epsilon_\Lambda/\Lambda$, where $\epsilon_\Lambda$ is the 
range of $\Lambda$-values that implement the desired cancellation, 
giving  
$\langle h\rangle^2\leq (246\ {\rm GeV})^2$ \cite{CS}. 
Once the uncertainty in 
$M_t$ is considered, the total fine-tuning is $\Delta\sim \int_\sigma 
d M_t\ \epsilon_\Lambda(M_t)/\int_\sigma  d M_t 
\Lambda=(1/\sigma)\int_\sigma d M_t\ \Delta_\Lambda(M_t)$, where $\sigma$ 
is the experimentally allowed 
range for $M_t$.}, 
and in this case this 
averaging has the effect of 
cutting the throat at $\Lambda\simeq 10$ TeV. This is characteristic of
fine-tuning arguments: since they are based on statistical
considerations, the conclusions may vary according to our partial
(and time-dependent) knowledge of the relevant parameters in the problem.
An alternative way of taking into account the uncertainty in $\Lambda$ 
and $M_t$ simultaneously is to add $\Delta_\Lambda$ and 
$\Delta_{\lambda_t}$ in quadrature (where the running top mass is 
$m_t=\lambda_t v/\sqrt{2}$):
\be
\label{Deltatot}
\Delta = \left(\Delta_\Lambda^2 + \Delta_{\lambda_t}^2\right)^{1/2}\
.  \ee
Since $\lambda_t$ should only vary within the experimentally allowed
range, we modify the definition of $\Delta_{\lambda_t}$ as
\be
\label{Deltat}
\Delta_{\lambda_t} = {\partial v^2\over \partial \lambda_t} {\lambda_t
\over v^2}\times {\delta^{\rm exp}\lambda_t\over \lambda_t} \ ,
\ee
(see ref.~\cite{CS} for a discussion of this point). In 
figure~\ref{fig:velt-c}
(left plot), which 
shows  the
$\Delta =$ 10, 100 curves, we indeed see a cut at $\Lambda\simeq 10$
TeV for $\Delta =$ 10. 
All this means that, in the absence of any theoretical reason
to be at Veltman's throat, or of precision measurements showing
that we are really there, fine-tuning arguments put an
upper limit on the value of $\Lambda_{\rm SM}$.

\begin{figure}[t]
\vspace{1.cm} \centerline{
\psfig{figure=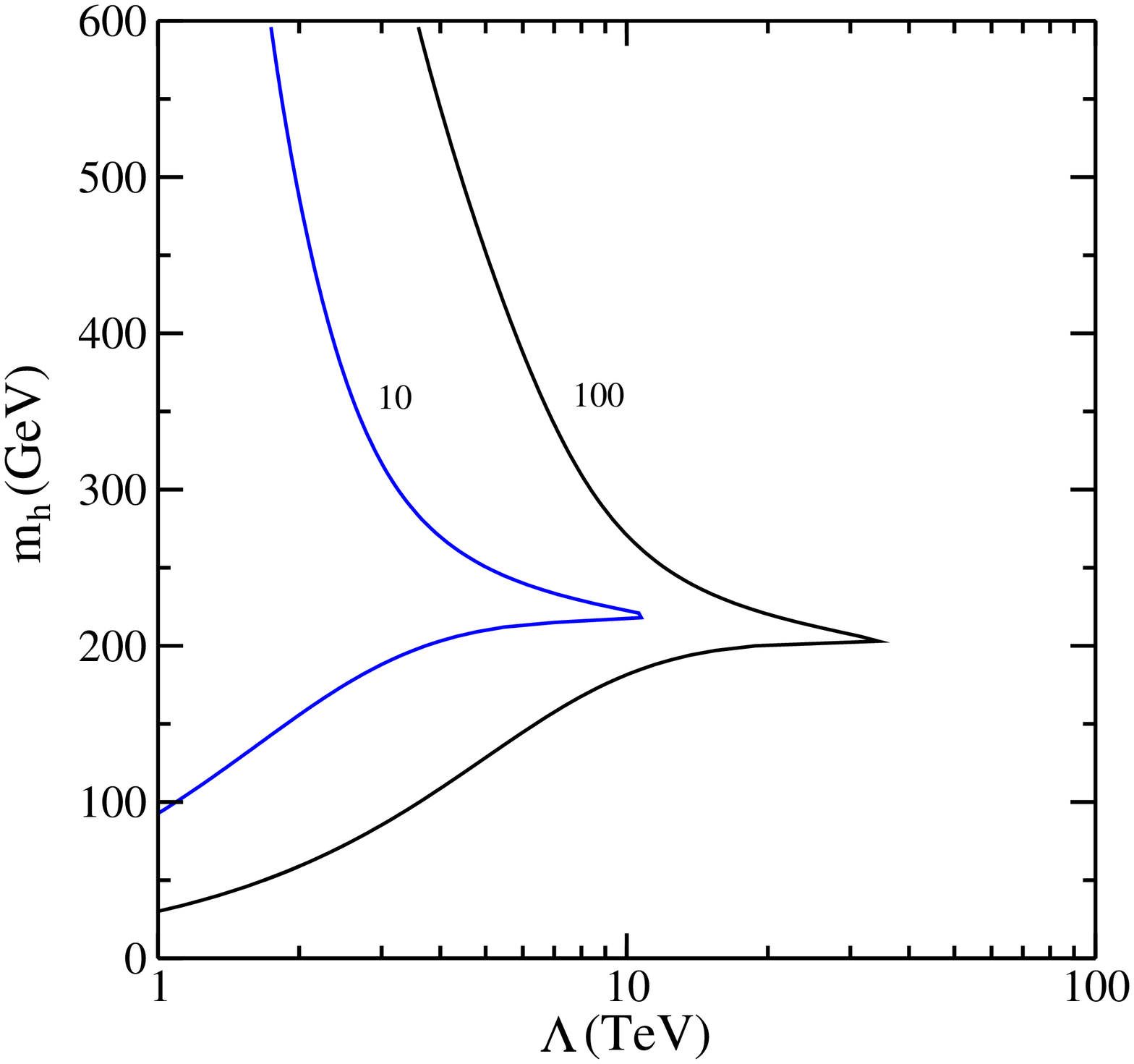,angle=-90,height=6cm,width=9cm,bbllx=4.cm,%
bblly=-3.cm,bburx=20.cm,bbury=20.cm}
\psfig{figure=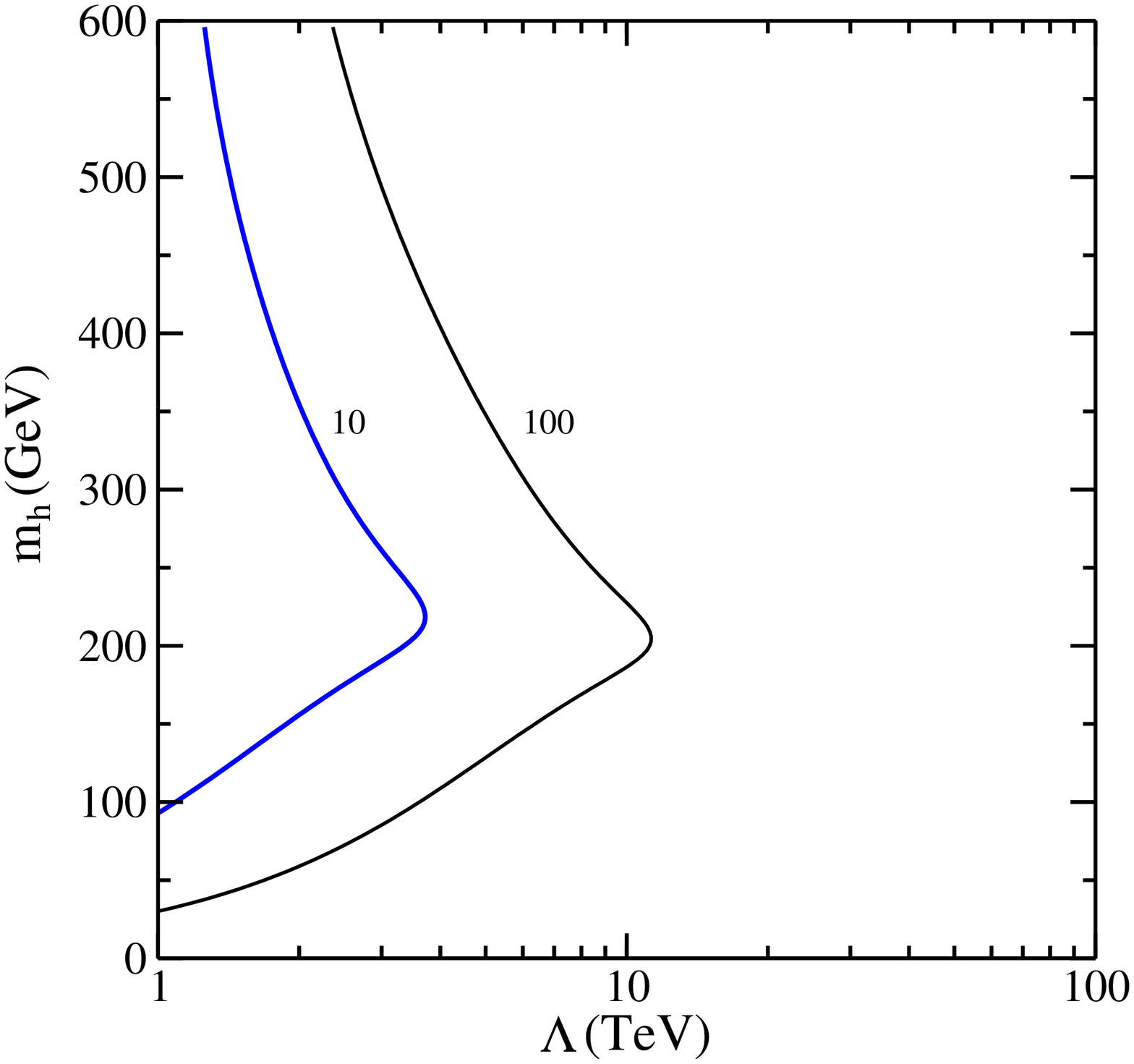,angle=-90,height=6cm,width=9cm,bbllx=4.cm,%
bblly=.5cm,bburx=20.cm,bbury=23.5cm}
}
\caption{\footnotesize
Contour plots of $\Delta=10, 100$. 
Left: with $\Delta$ 
as defined in eqs.~(\ref{Deltatot}, \ref{Deltat}).
Right: with $\Delta$ as defined in
eq.~(\ref{Deltatot2}).
}
\label{fig:velt-c}
\end{figure}

Our second remark concerns the fact that $m_h$ itself (and thus the
quartic Higgs coupling, $\lambda$) is not known at present. Therefore the
previous results, in particular the left plot in fig.~\ref{fig:velt-c},
correspond to a future time when $m_h$ will be known. For instance, if
LEP's inconclusive evidence for $m_h\simeq 115$ GeV \cite{LEPh} gets
confirmed, then one expects $\Lambda_{\rm SM}\simlt 1.4$ TeV. At 
present
one should average all the possible values of $m_h^2$, say in the range
115 GeV $\leq m_h \leq$ 600 GeV, which gives the result $\Lambda_{\rm
SM}\simlt 2.5$ TeV. Again, we can alternatively add in quadrature
$\Delta_{\lambda}$ to obtain the global fine-tuning,
\be
\label{Deltatot2}
\Delta = \left(\Delta_\Lambda^2 + \Delta_{\lambda_t}^2 +
\Delta_{\lambda}^2\right)^{1/2}\ . \ee
The corresponding curve for $\Delta =$ 10 is shown in 
the right plot of fig.~\ref{fig:velt-c}, which
corresponds to the present status of the problem.  
$\Lambda_{\rm SM}$ depends slightly on $m_h$, being always
below 4 TeV. In average,  $\Lambda_{\rm
SM}^{\rm av}\simlt 2.5$ TeV.

\section{Limitations of the use of the hierarchy problem to estimate 
${\bma \Lambda}_{\rm {\bf SM}}$}

Besides the previous subtleties about the shape of Veltman's throat,
there are more general caveats about using the
hierarchy problem to estimate the scale of new physics. The argument, as
usually presented and summarized in the Introduction, implicitly assumes 
that the job of the new physics is to cancel the dangerous contributions 
of the
SM diagrams for momenta above $\Lambda_{\rm SM}$, leaving
uncancelled the  quadratically divergent contributions evaluated 
in the SM and cut off at $\Lambda_{\rm SM}$
(except for artificial tunings or fortunate accidents).
However, the effects of the new physics do not enter in such
an abrupt and sharp way. Generically, the new diagrams
give a non-negligible contribution already below $\Lambda_{\rm SM}$,
and do not cancel exactly the SM contributions above $\Lambda_{\rm SM}$ 
\footnote{For related criticisms to the use of effective Lagrangians to 
estimate the effects of new physics see \cite{abuse}.}.
Remnants of this imperfect cancellation are finite and logarithmic 
contributions from NP, which are not simply given by the SM divergent part 
cut off at $\Lambda_{\rm SM}$. The familiar example is
SUSY, where the cancellation of the quadratically
divergent  contributions between the SM and the NP particles
automatically occurs for all momenta, even after the breaking of
SUSY.\footnote{From the point of view of the effective theory, the NP
contributions are threshold effects that are absorbed in the tree-level 
Higgs mass parameter of the effective Lagrangian with just the
right size to implement the cancellation.} 
If the only problem were this type of contributions, the
scale of SUSY breaking (and thus $\Lambda_{\rm SM}$) could be as large
as desired. Of course, this is not so because of the presence of other
dangerous logarithmic and finite contributions to the Higgs mass 
parameter, $m^2$, from superpartners. 
For instance, the stop 
sector,
$\tilde t$, contributes
\be
\label{stops}
\delta  m^2 = -{3 \lambda_t^2\over 8\pi^2}m_{\tilde t}^2  
\log {\Lambda^2\over m_{\tilde t}^2} \ .
\ee
This contribution does not cancel if SUSY is broken, setting finally
the bound  $\Lambda_{\rm SM}\simeq m_{\tilde t}\simlt$ few hundred GeV.
\footnote{This bound is actually stronger than suggested by the simplest 
argument based on the size of the quadratically-divergent contributions,
i.e. $\Lambda_{\rm SM}\simlt 2-3$ TeV. 
The reasons will be discussed in sect.4.}

The supersymmetric example clearly shows that Veltman's condition may be
irrelevant for estimating $\Lambda_{\rm SM}$. In SUSY the
quadratically-divergent contributions to $m^2$ are cancelled anyway, 
with or without Veltman's condition; but there are dangerous
contributions from new physics, which do not cancel (in principle), and
are totally unrelated to Veltman's condition.

The previous discussion can be generalized in a straightforward way.
For this matter, it is convenient to write the general 
one-loop effective potential using a momentum cut-off regularization,
$V=V_0 + V_1 + \cdots$ , with
\be
\label{V0}
V_0={1 \over 2} m^2 h^2 + {1 \over 4} \lambda h^4 \ ,
\ee
\be
\label{Vgen2}
V_1 = {1\over 64 \pi^2} {\rm Str}\left[
2\Lambda^2{\cal M}^2 + {\cal M}^4 \left(\log {{\cal M}^2\over \Lambda^2} 
- {1\over 2}\right)
\right] + {\cal O}\left({{\cal M}^6 \over \Lambda^2}\right)\ ,
\ee
where  $h$ is the (real and neutral) Higgs field,
the supertrace ${\rm Str}$ counts degrees of freedom with a
minus sign for fermions,
and ${\cal M}^2$ is the (tree-level, $h-$dependent) 
mass-squared matrix. The one-loop contribution to the Higgs mass
parameter is
\bea
\hspace{-1cm}
\delta_1 m^2 &=& 
\left. {\partial^2 V_1\over \partial h^2}\right|_{h=0} 
\nonumber
\\
&=&
{1\over 32 \pi^2} {\rm Str}\left[{\partial^2 {\cal M}^2\over \partial 
h^2}
\left(
\Lambda^2 + {\cal M}^2 \log {{\cal M}^2\over 
\Lambda^2}\right)
+ \left({\partial {\cal M}^2\over \partial h}\right)^2
\left( \log {{\cal M}^2\over \Lambda^2} + 1 
\right)\right]_{h=0}
\hspace{-0.5cm}.
\label{deltam1l}
\eea
%

\begin{figure}[t]
\vspace{1.cm}
\psfig{figure=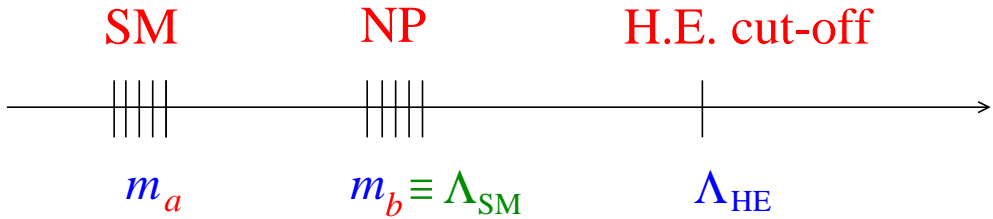,angle=-90,height=4cm,width=6cm,bbllx=8.cm,%
bblly=3.cm,bburx=14.cm,bbury=12.cm}
\caption{\footnotesize
Schematic representation of the spectra associated to 
the various scales of the theory: Standard Model (SM), 
New Physics (NP) beyond the Standard Model and the high energy 
cut-off.}
\label{fig:scheme}
\end{figure}
                                                                              
We will now separate the supertraces as sums over SM and NP 
states. The masses of the (lightest) NP states act
as the effective SM cut-off, $\Lambda_{\rm SM}$, but it is convenient
to consider the possible existence of a High-Energy cut-off \footnote{
This set-up can change if, above $\Lambda_{\rm 
SM}$, the Higgs shows up
as a composite field and/or new space-time dimensions open up.
However one might expect that the new degrees of freedom would play
a similar role as the NP states considered here, so that the
conclusions might not differ substantially.}, $\Lambda_{\rm HE}$, since 
the NP may not be yet the truly fundamental theory. This is schematically 
represented in fig.~\ref{fig:scheme}. 
$\Lambda_{\rm HE}$ can be as large as the Planck scale or much 
lower, close to the masses of the NP states.
Then
\bea
\hspace{-1cm}
\delta_1 m^2 &=&
{1\over 32 \pi^2} \sum_a^{\small \rm SM}N_a \left[ {\partial^2 
m_a^2\over \partial h^2}
\left(
\Lambda^2 + m_a^2 \log {m_a^2\over \Lambda^2}\right)
\right]_{h=0}
\nonumber\\
&+& 
{1\over 32 \pi^2} \sum_b^{\small \rm NP}N_b \left[ {\partial^2 
m_b^2\over \partial h^2}
\left(
\Lambda^2 + m_b^2 \log {m_b^2\over \Lambda^2}\right)+
\left({\partial m_b^2\over \partial h}\right)^2
\left( \log {m_b^2\over \Lambda^2} + 1 
\right)\right]_{h=0}
\hspace{-0.5cm},
\label{deltam1l2}
\eea
where $m_a, N_a$ ($m_b, N_b$) represent the mass and multiplicity, 
with negative sign for fermions, of the SM (NP) states, and
$\Lambda\equiv\Lambda_{\rm HE}$.
From the SM contributions, only the quadratically divergent ones are 
dangerous. The other terms are vanishing, except for the contribution
of the Higgs field itself, which is not large. On the other hand, all
NP contributions (quadratic, logarithmic and finite) are potentially 
dangerous. Now several situations might take place:

\begin{description}

\item[{\em i)}]
There are no special cancellations among the different contributions
in eq.~(\ref{deltam1l2}). In that case the simple (Big Hierarchy) 
argument, 
based on the size of the quadratic contributions, 
and the corresponding bound 
$\Lambda\simlt 2-3$ TeV, apply. The argument is clearly a conservative
one due to the presence of extra contributions, which are discussed below.


\item[{\em ii)}]
The SM quadratically divergent contributions cancel (maybe approximately)
by themselves, i.e. they are close to Veltman's condition. 
This is the situation discussed in the previous section.
There we saw that, in the absence of a fundamental reason for the exact
 cancellation, one expects NP not far from the TeV scale (``Veltman's 
throat" is always
cut). But, clearly the new states re-introduce the problem
[second line
of eq.~(\ref{deltam1l2})]: 
even if these new states do not give 
new quadratically divergent contributions, 
fine-tuning considerations require also $m_b\equiv\Lambda_{\rm SM}
\simlt 2-3$ TeV, as discussed next.

\item[{\em iii)}]
The SM and NP quadratically divergent contributions cancel each other, 
i.e.
\bea
\label{quadcancel}
\sum_a^{\small \rm SM}N_a  \left.{\partial^2 m_a^2\over \partial h^2}
\right|_{h=0}  + \sum_b^{\small \rm NP}N_b \left.{\partial^2 
m_b^2\over \partial h^2}
\right|_{h=0}=0\ .
\eea
This occurs naturally in SUSY and little-Higgs models.
But, again, the logarithmic and finite
NP contributions in eq.~(\ref{deltam1l2}) are also dangerous and, as in 
case {\em ii)}, re-introduce the problem, albeit in a softer form. 
Notice also that these contributions (unlike the quadratic ones)
show up in any regularization scheme, differing only in the
value of the finite pieces. For future use, we write them in the 
$\overline{\rm MS}$ 
scheme\footnote{For simplicity we are using the $\overline{\rm DR}$ 
variation of the $\overline{\rm MS}$ scheme, so that all finite
pieces in (\ref{deltam1MS}) have the same coefficient, independently of the
nature of the $b-$states.}
\be
\label{deltam1MS}
\delta_{\rm NP}^{\overline{\rm MS}}\ m^2 =
 \sum_b^{\small \rm NP}
{N_b\over 32 \pi^2}
\left[ {\partial^2 
m_b^2\over \partial h^2}
m_b^2\left( \log {m_b^2\over Q^2}-1\right)+
\left({\partial m_b^2\over \partial h}\right)^2
\log {m_b^2\over Q^2} \right]_{h=0}\ ,
\ee
where $Q$ is the renormalization scale, to be identified with the 
cut-off scale, $\Lambda$.
Quantitatively, these contributions are roughly similar to the 
SM quadratically-divergent one, replacing 
$\Lambda\rightarrow m_b\equiv\Lambda_{\rm SM}$. This gives the basis for 
the estimate of the ``naive" Big Hierarchy argument
discussed in the Introduction.
However, new parameters not present in the SM might enter through the 
$m_b$ masses and, moreover, the presence of the 
logarithmically-enhanced terms 
makes the new contributions typically more dangerous 
than the SM estimate (as happens 
for instance in the supersymmetric case commented above). 
Hence,
from fine-tuning arguments we can keep
$m_b\equiv\Lambda_{\rm SM}
\simlt 2-3$ TeV, as a conservative bound.

Of course, if the NP is itself an effective theory
derived from a more fundamental one, further extra states 
[which could have ${\cal O}(M_p)$ masses] would be even more
dangerous, unless their contributions are under control for some reason. 
The only clear example of this desirable property occurs when the theory 
is supersymmetric.

\item[{\em iv)}]
It may happen that, besides the cancellation of quadratic
contributions, the other dangerous
contributions also cancel or are absent.
In the $\overline{\rm MS}$ scheme this means that
eq.~(\ref{deltam1MS}) vanishes.
This could happen by accident or for some fundamental reason. In either
case, the scale of new physics could be really much larger than 
a few TeV. The obvious remark is that no such fundamental reason
is known.
In its absence, 
one has to average 
over the possible ranges of variation of the parameters defining
the new physics (e.g. the soft masses for the MSSM). In this way
the usual result
$\Lambda_{\rm SM}\simlt 2-3$ TeV is generically recovered.

To be on the safe side, however, one should not disregard completely the 
possibility
that eq.~(\ref{deltam1MS}) could be vanishing, which would allow for 
a very large $\Lambda_{\rm SM}$. It is fun to imagine some
scenarios of this kind. E.g. if the NP states get masses through
conventional Yukawa couplings (this is the case of a fourth
generation), eq.~(\ref{deltam1MS}) vanishes. 
Notice that, strictly speaking, those extra states do not represent a genuine
new scale of physics as their masses arise from the electroweak
breaking scale. In any case, the cancellation
of the quadratic divergences would require to be close to a (modified)
Veltman's condition, and the discussion of point {\em ii)} above would follow.
In addition, a heavy fourth generation would have problems of
perturbativity and stability of
the Higgs potential, among others. Another possibility is that
the SM is close to Veltman's condition and the new physics corresponds
to heavy supersymmetric representations [possibly plus 
${\cal O}({\rm TeV})$ soft breaking terms] (such 
partially-supersymmetric models have been already considered in the 
context of extra dimensions \cite{Alex}). A further possibility will be 
discussed at the end of sect.~4.

Although the previous examples do not seem realistic, it is not 
unconceivable that the unknown fundamental physics is smart enough
to implement naturally a cancellation of eq.~(\ref{deltam1MS}),
as SUSY does with the quadratic divergences. In that case fine-tuning
arguments would be misleading.

\end{description}

\noindent
Leaving aside the last caveats, it follows from the previous discussion
that the ``naive" procedure to estimate the scale of NP, 
by considering the SM quadratically-divergent contribution with
$\Lambda\rightarrow \Lambda_{\rm SM}$, works reasonably well in most
cases. Actually it is a 
conservative one, since it disregards unknown contributions that tend to 
be larger.
On the other hand, it is not reliable to make more detailed statements 
based on this argument.
In particular, the conditions for the cancellation (or quasi-cancellation)
of the SM dangerous contributions are in principle unrelated
to the cancellation of the unknown dangerous contributions.
At this point the effective theory is unable to provide information
about the dangers of the new physics.

Therefore, to derive more accurate implications for new physics from 
fine-tuning arguments, one should consider specific possibilities
for physics beyond the SM. We examine two relevant examples in the rest of 
the paper.

\section{Examples of New Physics}

\subsection{Right-handed seesaw neutrinos}

The simplest extension of the ordinary
SM is obtained by adding right-handed neutrinos, $\nu_R$ (one per 
family).
The usual goal is to describe  neutrino masses, either by
conventional Yukawa couplings or through a seesaw mechanism.
The latter is probably the most elegant mechanism to
explain the smallness of neutrino masses. Then, the SM Lagrangian
is simply enlarged with
\bea
\label{Lss}
{\cal L}_\nu= -\nu_R m_D \nu_L - {1\over 2} \nu_R M_R  \nu_R^T 
+ {\rm h.c.}\ ,
\eea
where $m_D$ is the Dirac mass matrix ($m_D= -\lambda_\nu h/\sqrt{2}$,
where $\lambda_\nu$ is the neutrino Yukawa coupling, a matrix in flavour 
space) and $M_R$ is the Majorana mass matrix
for right-handed neutrinos. For our purposes it is enough to consider a single
neutrino species, so that the above matrices reduce to numbers. 
For $M_R^2\gg m_D^2$, the two mass-squared eigenvalues are 
\bea
\label{masseigen}
m_{\nu_1}^2&=&{m_D^4\over M_R^2} + {\cal O}\left({m_D^6\over M_R^4} 
\right)\ ,
\nonumber\\
m_{\nu_2}^2&=&M_R^2 +2m_D^2 + {\cal O}\left({m_D^4\over M_R^2} \right)\ .
\eea
The lightest eigenstate is part of the effective low-energy theory and
does not contribute radiatively to the Higgs mass parameter,
as it is apparent from eq.~(\ref{deltam1l2}). 
The heaviest one, however,
contributes not only to the quadratic divergence, but also to the finite 
and logarithmic parts of $\delta_1 m^2$:
\bea
\label{deltanu}
\delta_\nu m^2=-{\lambda_\nu^2\over 16\pi^2}
\left[2\Lambda^2 + 2M_R^2 \log{M_R^2\over \Lambda^2}\right]\ .
\eea
This equation is a particularly simple example of the second line of
eq.~(\ref{deltam1l2}) and illustrates the general discussion 
of the previous section. In particular, even if the quadratically
divergent contributions of the SM cancel the one  of 
eq.~(\ref{deltanu}) (which, incidentally, means that Veltman's
condition is modified by undetectable physics), there are other dangerous 
contributions which do not cancel. 
In other words, this represents a new manifestation
of the hierarchy 
problem \cite{CCIQ} since a cancellation between the quadratic and the
logarithmic and finite contributions would be completely artificial
and depends on the choice of renormalization scheme.

The logarithmic and finite contributions are especially
disturbing as their size is associated to $M_R$ (which is expected
to be very large) and show up in any renormalization scheme.
In the $\overline{\rm MS}$ scheme [see eq.~(\ref{deltam1MS})]
\bea
\label{deltanuMS}
\delta_\nu^{\overline{\rm MS}} m^2=-{\lambda_\nu^2\over 8\pi^2}
M_R^2 \left[\log{M_R^2\over \Lambda^2}-1\right]\ ,
\eea
where we have already set $Q=\Lambda$.
Now it is easy to obtain a lower bound on the size of $M_R$ from
the request of no fine-tuning. Demanding
\bea
\label{boundnu1}
\left|{\delta_\nu m^2\over m^2}\right|\leq \Delta\ ,
\eea
requires
\be
\label{boundnu2}
M_R< 10^7\ {\rm GeV}\ \Delta^{1/3}
\left({m_h\over 200\ {\rm GeV}}\right)^{2/3}
\left({m_\nu\over 5\times 10^{-2}\ {\rm eV}}\right)^{-1/3}
\left[\log{\Lambda^2\over M_R^2} + 1\right]^{-1/3}\ .
\ee
Hence, for any sensible value of $\Delta$, we obtain a quite robust bound 
\bea
\label{boundnu3}
M_R\simlt 10^7\ {\rm GeV}\ ,
\eea
which can only be satisfied if the neutrino Yukawa coupling is
very small, $\lambda_\nu \simlt 10^{-4}$. This is possible,
but spoils the naturalness of the seesaw mechanism to explain
the smallness of $m_\nu$.

Of course, the previous bound has been obtained under the assumption
that the only NP are the right-handed neutrinos. In a SUSY scenario,
the contribution of their scalar partners (sneutrinos) render $\delta_\nu 
m^2$
small and not dangerous. The obvious conclusion is that,
in the context of the SM, the seesaw mechanism suffers a very 
important 
fine-tuning problem which cannot be evaded by 
invoking e.g. a Veltman-like cancellation of the quadratically
divergent contributions to $m^2$.
Thus the seesaw mechanism claims for the existence of additional NP.
In this sense, SUSY seems to be the favourite framework to
accommodate it.

\subsection{SUSY}

SUSY is the paradigmatic example of a theory 
in which the quadratically divergent radiative
corrections of the SM are cancelled by those coming from new physics,
beautifully realizing eq.~(\ref{quadcancel}). The new particles introduced
by SUSY (gauginos, sfermions and higgsinos) have masses of the order of
the scale of the soft breaking terms, ${\cal O}(m_{\rm soft})$. In the
scheme of fig.~\ref{fig:scheme}, $m_b \sim m_{\rm soft}$.  
According to the general discussion in
point {\em iii)} of sect.~3, despite the cancellation of the quadratic
corrections, logarithmic and finite contributions due to
the new states lead to the usual (but conservative) fine-tuning upper 
bound $m_b\equiv\Lambda_{\rm SM} \simlt 2-3$ TeV.

In fact, according to the usual analyses, in the Minimal Supersymmetric
Standard Model (MSSM), the absence of fine tuning requires a more
stringent upper bound, namely $m_{\rm soft}\simlt$ few hundred GeV
\cite{BG,CS,ftgalore,CEH}. Actually, the available experimental 
data already imply that typically the
ordinary MSSM is fine-tuned at the few percent level.  The reasons for
this abnormally acute tuning of the MSSM have been reviewed in
refs.~\cite{KM,CEH} (for related work see 
refs.~\cite{BG,CS,ftgalore}),
and we summarize and update them here for the sake of completeness.

\vspace{0.3cm}
\noindent
{\em MSSM}

\noindent
In the MSSM the Higgs 
sector
consists of two $SU(2)_L$ doublets, $H_1$, $H_2$. The 
(tree-level) scalar potential for their neutral components, $H^0_{1,2}$, 
reads
\be
\label{VMSSM}
V^{\rm MSSM}=m_1^2|H_1^0|^2+m_2^2|H_2^0|^2- (m_3^2
H_1^0 H^0_2 + {\rm h.c.})+ \frac{1}{8}(g^2+g'^2) 
(|H^0_1|^2-|H^0_2|^2)^2,
\ee
with $m_{1,2}^2=\mu^2 + m_{H_{1,2}}^2$ and $m_3^2=B\mu$, where 
$m_{H_{i}}^2$ and $B$ are soft masses and $\mu$ is the
Higgs mass term in the superpotential, $W\supset \mu H_1\cdot  H_2$.
Minimization of $V^{\rm MSSM}$ leads to VEVs for $H_1^0$ and $H_2^0$.
Along the breaking direction in the $H_1^0, H_2^0$ space, 
$h=\sqrt{2}\left(\cos \beta\ {\rm Re}H_1^0 
+ \sin \beta\ {\rm Re}H_2^0\right)$ (with $\tan\beta\equiv \langle 
H^0_2 \rangle/\langle H^0_1 \rangle$),
the potential 
(\ref{VMSSM}) can be written in a 
SM-like form:
\be
\label{Vbeta}
V={1 \over 2} m^2 h^2 + {1 \over 4} \lambda h^4 \ ,
\ee
where $\lambda$ and $m^2$ are functions of the 
initial parameters, which for the MSSM are the 
soft masses and the $\mu-$parameter at the initial (high energy) 
scale. Note in particular that $m^2$ contains at tree-level
contributions of order 
$m_{\rm soft}^2$ and $\mu^2$.
Minimization of (\ref{Vbeta}) gives the usual SM-like result
\be
v^2=\langle h^2 \rangle={-m^2\over \lambda} \ .
\ee

Now, $m^2$ contains contributions $\sim  m_{\rm soft}^2$ already at tree-level.
So, roughly speaking, $\Delta\simeq m_{\rm soft}^2/(\lambda v^2)$ and
the absence of fine-tuning at the 10\% level
requires $m_{\rm soft}^2 \simlt 10 m^2\simeq 5 m_h^2$, i.e. 
$m_{\rm soft} \simlt$ few hundred GeV. Notice also that the larger 
the tree-level value of $m_h$ the 
smaller the fine-tuning, which is a very generic fact.

Since the soft masses do not need to be degenerate, it could happen that 
those of the Higgses are smaller than e.g. those of the squarks, and so  
the previous bound does not translate automatically into a bound
on squark masses. However, squark masses enter in $m^2$ through
logarithmic and finite loop corrections [given by 
eq.~(\ref{deltam1l2})]. In particular, the contribution from the stop sector
 has already been written in (\ref{stops}), where it is clear that the
loop factor is largely compensated by the multiplicity of states, the
sizable top Yukawa coupling and the logarithmic factor\footnote{Actually,
these effects are needed for the usual mechanism of radiative breaking of
$SU(2)\times U(1)$, since they naturally provide a negative contribution
to $m^2$.}. These corrections can also be viewed, and computed more 
precisely, as the effect of the RG running of $m^2$ from the high scale
down to the electroweak scale. The consequence is that the absence of
fine-tuning puts upper bounds on the soft masses, which are even stronger
than those from the previous ``tree-level" argument. Namely, at the
initial high-energy scale, all soft masses (at least those of the
third-generation sfermions and the gluino) are confined to the $\simlt 
200$
GeV region in order to avoid a fine-tuning larger than 10\%.

This situation is worsened by the smallness of the quartic Higgs 
coupling. At tree-level
$\lambda =(1/8)(g^2+g'^2)\cos^2 2\beta$  which sets the upper bound
$m_h^2\simeq 2\lambda v^2\leq M_Z^2\cos^2 2\beta$. 
The value of $\lambda$ (and thus $m_h$) increases due to radiative 
corrections, and this reduces in principle the
fine-tuning. These corrections, which depend also on the soft masses as
$\delta_{\rm rad}\lambda\propto\log(m_{\rm soft}^2/m_t^2)$, modify
the theoretical upper bound on $m_h$ in the MSSM as
\bea
\label{mhMSSM}
m_h^2\leq M_Z^2 \cos^2 2 \beta + {3m_t^4 \over 2\pi^2 v^2}
\log{M_{\rm SUSY}^2\over m_t^2} + ...  
\eea
where $m_t$ is the (running) top mass and $M_{\rm SUSY}\propto 
m_{\rm soft}$ is an average of stop masses. These radiative corrections
are indeed mandatory in order to increase
$m_h$ beyond the  experimental bound, $m_h\geq 
115$ GeV.
The problem is that a given increase in $m_{\rm soft}^2$ reflects linearly 
in the partial contributions to ${m}^2$ and only logarithmically in $\lambda$
(and $m_h^2$), 
so the fine tuning $\Delta\sim 2m_{\rm soft}^2/m_h^2$ gets usually worse. 
Since these corrections are mandatory, one is forced to live in a
region of relatively large soft masses, $m_{\rm soft}\geq 300$ GeV,
which implies a substantial fine-tuning, $\Delta \simgt 20$  
\cite{BG,CS,ftgalore,CEH}.

It is interesting to note that the recent shift in the measured top mass, 
as quoted in eq.~(\ref{topmass}), has a positive impact
(modest but not negligible) on the MSSM
fine-tuning. The 4 GeV increase of the central value of $M_t$ (from 174
GeV to 178 GeV) implies that the radiative corrections to the Higgs mass
in (\ref{mhMSSM}) are larger, so that stop soft masses can be 
smaller, 
thus 
reducing the fine tuning. It is not difficult to estimate the improvement
in $\Delta$ from eq.~(\ref{mhMSSM}). For a given value of $m_h$ a
(positive) shift of $m_t$ reflects on a (negative) shift of $M_{\rm
SUSY}^2$:
\bea
\label{variations}
{\delta M_{\rm SUSY}^2\over M_{\rm SUSY}^2}\simeq
-2{\delta m_t\over m_t}\left[10^{-3}\ {\rm GeV}^{-2}
(m_h^2 - M_Z^2 \cos^2 2 \beta) -1 \right] \simeq - 
8\  {\delta m_t\over m_t}\ ,
\eea
where the last figure corresponds to large $\tan\beta$ ($\geq 8$)
and $m_h=115$ GeV,  
which is the most favorable case for the fine-tuning
in the MSSM. Since the value of $\Delta$ is approximately proportional
to $m_{\rm soft}^2$ (for more details, see the discussion
around eq.~(15) of ref.~\cite{CEH}), we conclude that the 4 GeV increase
of the experimental value of the top mass leads to a 20\% decrease of 
$\Delta$.
A former typical value $\Delta = 35 \pm 9$ for $M_t=174.3\pm 5.1$ GeV 
becomes now $\Delta = 28 \pm 7$ for $M_t=178\pm 4.3$ GeV.
(A numerical calculation of $\Delta$ confirms the accuracy of this
estimate.)

\vspace{0.3cm}
\noindent
{\em Other scenarios}

\noindent
As discussed above, the fine tuning of the MSSM is much more severe than
naively expected due, basically, to the smallness of the tree-level Higgs
quartic coupling, $\lambda_{\rm tree}$ and, also, to the large magnitude
of the RG effects. Any departure from the ordinary MSSM that improves
these aspects will also potentially improve the fine-tuning situation. In
particular, a larger $\lambda_{\rm tree}$ can be obtained in models with
extra dimensions opening up not far from the electroweak scale
\cite{Strumia}, or by extending either the gauge sector \cite{extG} or the
Higgs sector \cite{extH} (as it happens in the NMSSM \cite{NMSSM}).  
Another possibility, which simultaneously increases $\lambda_{\rm tree}$
and reduces the magnitude of the RG effects, is to consider scenarios in
which the breaking of SUSY occurs at a low scale (not far from the TeV
scale)  \cite{hard,Brignole,Polonsky,BCEN}. One may wonder if it is
possible in any of these scenarios to saturate the generic fine-tuning
upper bound $\Lambda_{\rm SM} \simlt 2-3$ TeV, discussed in sect. 3. This
is actually the case, at least in the latter scenario, which we briefly
discuss next.

There are two scales involved in the breaking of SUSY (\SUSY): $\sqrt{F}$,
which corresponds to the VEVs of the relevant auxiliary fields in the
\SUSY sector; and the messenger scale, $M$, associated to the
non-renormalizable interactions that transmit the breaking to the
observable sector. These non-renormalizable operators give rise to soft
terms (such as scalar soft masses) and also hard terms (such as quartic
scalar couplings): 
\be
\label{mlambda}
m_{\rm soft}^2\sim {F^2\over M^2}\ ,\;\;\;\; \lambda_{\tinySUSY}\sim
{F^2\over M^4}\sim {m_{\rm soft}^2\over M^2}\ .
\ee
Naturalness requires $m_{\rm soft} \leq {\cal O}(1\, {\rm TeV})$, but
this does not fix the scales $\sqrt{F}$ and $M$ separately. So, unlike in
the MSSM, the scales $\sqrt{F}$ and $M$ could well be of similar order
(thus not far from the TeV scale), which corresponds to low-scale \SUSY 
scenarios \cite{hard,Brignole,Polonsky,BCEN}.  In this
framework, the hard terms of eq.~(\ref{mlambda}), are not negligible and
the \SUSY contributions to the Higgs quartic coupling can be easily
larger than the ordinary MSSM value. As a consequence, the tree-level
Higgs mass can be much larger than in the MSSM, radiative corrections are
not needed, and the scenario can be easily consistent with experimental
bounds with virtually no fine-tuning.

The value of $\Lambda_{\rm SM}$ is given by the masses of the extra
particles. As for the MSSM, upper bounds on these masses are obtained by
requiring that the tree-level and the radiative contributions to $m^2$ are
not too large, in order to avoid fine-tuning. Again as in the MSSM, the
tree-level contributions are generically ${\cal O}(m^2_{\rm soft})$, so 
that the absence of
fine-tuning requires $m_{\rm soft}\simlt$ few hundred GeV.

 This seems to
be a {\em very generic} fact of any SUSY scenario, but it only affects the
(extended) Higgs sector. In other words, fine-tuning considerations imply
that in a generic SUSY framework there are new states from the Higgs 
sector below 1~TeV. Actually, the argument extends to Higgsinos, as they 
get
masses ${\cal O}(\mu)$, which should be of the same order as the Higgs
soft masses (note that a significant mixing with gauginos only occurs if
the gaugino soft masses are also small).

On the other hand, bounds on squark and gluino masses come from the
radiative corrections to $m^2$. In the low scale \SUSY framework, these
contributions are much smaller than in the MSSM, since the logarithmic
factors are ${\cal O}(1)$. E.g. from the stop contribution written in
eq.~(\ref{stops}), a 10\% fine-tuning in the Higgs mass requires
$m_{\tilde t}\simlt 7-10 m_h$, which for $m_h =250-350$ GeV saturates the
bound $\Lambda_{\rm SM} \simlt 2-3$ TeV. These values for the Higgs mass
may not be favoured by the present electroweak data, but in any case 
they can be easily reached in the framework of low-scale \SUSY \cite{CEH}.

\vspace{0.3cm}
\noindent
{\em A peculiar SUSY scenario}

\noindent
We have seen how in generic SUSY models the usual fine-tuning bound
$\Lambda_{\rm SM} \simlt 2-3$ TeV holds, although in many cases the bound
is more stringent due to finite and logarithmic contributions to $m^2$,
which have no reason to cancel. However, it is amusing to think of a
scenario of the type {\em iv)} discussed in section~3, where this 
additional cancellation takes place, i.e. where
eq.~(\ref{deltam1MS}) vanishes.

If a non-accidental cancellation occurs in (\ref{deltam1MS}), a
plausible possibility seems to require universality, {\it i.e.}
$m_b(h=0)\equiv \tilde m$, for all particles beyond the SM 
ones. This universality does not exactly coincide 
with the universality of the soft breaking terms usually 
invoked in MSSM analyses. The differences occur in the Higgs/higgsino 
sector: degeneracy of higgsinos with the other states requires 
adjusting also  the $\mu$ parameter. Besides, now the Higgs soft masses 
are
not equal to the other soft masses; instead, they have to be adjusted 
so that one Higgs doublet (some combination of $H_1$ and $H_2$) is heavy 
and degenerate with the other states while the orthogonal combination is
kept light and plays the role of the SM Higgs.

We will be interested in considering the possibility of 
$\tilde m\gg m_W$ even if this is against the usual naturalness argument 
for bounding the soft masses in the MSSM. In this universal case, 
eq.~(\ref{deltam1MS}) reads
\be
\delta_{\rm NP}^{\overline{\rm MS}}\ m^2 ={1\over 32 \pi^2}
\sum_b^{\small \rm NP}N_b
\left[ \tilde m^2\left( \log {\tilde m^2\over Q^2}-1\right)
\left.{\partial^2 m_b^2\over \partial h^2}\right|_{h=0}
+\log {\tilde m^2\over Q^2}
\left.\left({\partial m_b^2\over \partial h}\right)^2\right|_{h=0}
\right]\ .
\ee
The logarithmic and finite contributions in this expression have a clear 
interpretation in the language of effective field theories. The logarithms 
can be interpreted as coming from RG running beyond the scale $\tilde m$, 
while the finite contributions can be interpreted as threshold corrections 
coming from integrating out the physics at $\tilde m$. In fact this 
threshold correction is just $\delta_{\rm NP}^{\overline{\rm MS}}\ 
m^2$  evaluated at the scale $Q=\tilde m$. In this section we will 
require only that this threshold correction vanishes and therefore we
disregard the possible effects from running beyond the scale $\tilde m$.
(In other words, we are imposing the universality condition at $Q=\tilde 
m$.) Needless to say, such effects might spoil the electroweak hierarchy 
(unless $\tilde m$ is the cut-off scale in the fundamental theory, see 
below) but we are being conservative and only require that the NP 
particles at $\tilde m$ do not destabilize that hierarchy themselves.
After setting then $Q=\tilde m$ we get
\be
\delta_{\rm NP}^{\overline{\rm MS}}\ m^2 (Q=\tilde m)=
-{\tilde m^2\over 32 
\pi^2}\sum_b^{\small \rm NP}N_b
\left.{\partial^2 m_b^2\over \partial h^2}\right|_{h=0}\ .
\label{sillysusy}
\ee
Using now the fact that the theory is supersymmetric and there is a 
cancellation of quadratic divergences, we can use 
(\ref{quadcancel}) to rewrite (\ref{sillysusy}) in the form
\be
\delta_{\rm NP}^{\overline{\rm MS}}\ m^2 (Q=\tilde m)=
{\tilde m^2\over 32 \pi^2}\sum_a^{\small \rm SM}N_a
\left.{\partial^2 m_a^2\over \partial h^2}\right|_{h=0}\ .
\label{sillysusy2}
\ee
This result is interesting because it involves only SM
particles\footnote{It is instructive to work out explicitly
(\ref{sillysusy}) to check (\ref{sillysusy2}). One has to make use of the
SUSY relation $\lambda=(1/8)(g^2+g'^2)\cos^2 2\beta$.} and therefore the
potentially large quantity $\delta_{\rm NP}^{\overline{\rm MS}}\ m^2
(Q=\tilde m)$ is proportional to the same combination of couplings that
appears in Veltman's condition. In this very peculiar scenario then,
imposing Veltman's condition would also work for the cancellation of the
dangerous contributions of NP particles.

The first concern is that Veltman's condition can be satisfied in 
the SM by adjusting the unknown Higgs mass,
while in the MSSM 
there are strong bounds on the latter. In fact, Veltman's ``prediction'', 
$m_h\simeq 307$ GeV seems to be hopelessly large for the MSSM. However, 
this not so: as we have seen, RG effects are important to evaluate a 
refined value of $m_h$ coming from Veltman's condition, 
\be
\label{V2}
3g^2 + g'^2 + 8\lambda -
8\lambda_t^2 = 0\ , 
\ee
which is supposed 
to hold at $\Lambda$. There are two competing effects. First, $\lambda_t$ 
gets 
smaller and smaller when the energy scale increases and consequently 
the predicted $\lambda(\Lambda)$ gets smaller for increasing $\Lambda$.
Second, for a given $\lambda(\Lambda)$ a larger interval of running 
causes $\lambda$, and thus $m_h$, to be bigger at the 
low scale. The first effect turns out to win and $m_h$ decreases with 
increasing $\Lambda$ (see figure~\ref{fig:velt-a}). 
In the peculiar SUSY scenario we are discussing, one should take these 
important effects into account to see whether Veltman's condition can be 
satisfied at some scale. Remarkably, the scale at which Veltman's 
condition holds [with $\lambda=(1/8)(g^2+g'^2)\cos^22\beta$] turns out to 
be around the string scale! More precisely, $\Lambda\simeq 10^{18}$ GeV 
for $\tan\beta\gg 1$ and  $\Lambda\simeq 10^{25}$ GeV for $\tan\beta=1$. 
Running $\lambda$ down in energy one 
obtains $m_h\simeq 140-150$ GeV as a further prediction of this model.

We consider this scenario, somewhat reminiscent of Split Supersymmetry
\cite{splitsusy}, as a mere curiosity. The reasons that prevent us from
taking it seriously are manifold: first, there is in principle no 
theoretical
reason to expect that Veltman's condition should be satisfied, even though
the couplings $g, g', \lambda$ and $\lambda_t$ can all be related to the
string coupling, and for a particular string vacuum this could be the
case. Second, the fulfillment of Veltman's condition at higher loop
order is more difficult to justify or even impossible (which is a 
problem given the large value of $\tilde m$). Third, the condition
$m_b^2(h=0)=\tilde m$ is equally difficult to justify theoretically,
especially in the Higgs sector, which involves both SUSY and soft masses.
Even generating $\mu$ through the Giudice-Masiero mechanism \cite{GM} 
requires tuning to achieve the desired universality. Finally, the MSSM 
relation $\lambda=1/8(g^2+g'^2)\cos^22\beta$ we have used to evaluate 
$\tilde m$ receives \SUSY contributions which (as discussed above) are 
important for $\sqrt{F}\sim M$, which is the case now. Moreover, the 
appropriate framework to study this problem is SUGRA rather than the 
conventional MSSM with global SUSY.

\section{Summary and Conclusions}

In the first part of this paper (sections 1--3) we have re-examined the
use of the Big Hierarchy Problem of the SM to estimate the scale of New
Physics (NP). The common argument is based on the size of the
quadratically-divergent contributions to the squared Higgs mass parameter,
$m^2$. Treating the SM as an effective theory valid below $\Lambda_{\rm
SM}$, and imposing that those contributions are not much larger than
$m_h^2\simeq 2 m^2$ itself, one obtains $\Lambda_{\rm SM} \simlt 2-3\ {\rm
TeV}$.

It has been argued in the literature (e.g.~\cite{KM}) that, if $m_h$ lies
(presumably by accident) close to the value that cancels the quadratic
contributions (i.e. the famous Veltman's condition), $\Lambda_{\rm SM}$
could be much larger. However, as we have shown in section~2, a complete
evaluation of the fine-tuning (which should include the sensitivity to the
top Yukawa coupling and the Higgs self-coupling) indicates that this is
not the case at present.

Then, we have examined the limitations of the Big Hierarchy argument to
estimate the scale of new physics. In our opinion the reasoning, as
usually presented and summarized above, is arguably too naive, as it
implicitly assumes that the SM quadratically divergent contributions, cut
off at $\Lambda_{\rm SM}$, remain uncancelled by the effect of NP, except
for accidents or artificial tunings. However, the NP diagrams give a
non-negligible contribution already below $\Lambda_{\rm SM}$, and do not
cancel exactly the SM contributions above $\Lambda_{\rm SM}$. Remnants of
this imperfect cancellation are finite and logarithmic contributions from
NP, which are not simply given by the SM divergent part cut off at
$\Lambda_{\rm SM}$ (the familiar example is SUSY, as discussed in
sect.~3). The general analysis presented here, based on a
model-independent study of the one-loop effective potential, shows that,
quantitatively, these contributions are typically larger than the estimate
of the ``naive" Big Hierarchy argument. Hence, from fine-tuning arguments
one can keep $\Lambda_{\rm SM} \simlt 2-3$ TeV as a {\em conservative}
bound.

Although very general, this kind of analysis still has limitations.  One
is that it assumes that the (4D) Higgs field continues to be a fundamental
degree of freedom above $\Lambda_{\rm SM}$. Another is that, besides the
cancellation of quadratic contributions, the other dangerous NP
contributions [shown explicitly in eq.~(\ref{deltam1MS})] could also
cancel or be absent. This might happen by accident or for some fundamental
and unknown reason (we have presented some amusing, though not very
realistic, examples). In that case fine-tuning arguments would be
misleading.

In order to derive more accurate implications for new physics from
fine-tuning arguments, one must consider specific possibilities for NP.  
This is the goal of the second part of the paper (section 4), where we
have examined in closer detail two physically relevant examples of NP:
right-handed (seesaw) neutrinos and SUSY. They also illustrate the general
conclusions obtained in the previous sections.

Right-handed seesaw neutrinos with a large Majorana mass, $M_R$,
contribute quadratically divergent corrections to the Higgs mass, as well
as finite and logarithmic contributions which are especially dangerous as
their size is associated to $M_R$ (expected to be very large).  Demanding
that these finite and logarithmic corrections are not much larger than
$m^2$ itself translates into the upper bound $M_R\simlt 10^7\ {\rm GeV}$,
which spoils the naturalness of the seesaw mechanism to explain the
smallness of $m_\nu$.  The conclusion is that, in the context of the SM,
the seesaw mechanism suffers a very important fine-tuning problem which
claims for the existence of {\em additional} NP. In this sense, we have
argued that SUSY is the favourite framework to accommodate the seesaw
mechanism.

Finally, SUSY is a typical example where the ``Big Hierarchy'' bound,
$\Lambda_{\rm SM} \simlt 2-3$ TeV, is indeed {\em conservative}: standard
fine-tuning analyses (updated in this paper to incorporate the upward
shift in the top mass) show that the MSSM is fine-tuned at least by a few
\% for soft masses $\simlt$ few hundred GeV. We have reviewed the reasons
for this abnormally acute tuning, an important one being the logarithmic 
enhancement of the NP corrections to $m^2$.

We have shown that the MSSM problems can be alleviated in other types of
scenarios. In particular, we have stressed the fact that low \SUSY
scenarios can easily evade the problematic MSSM aspects, decreasing the
fine-tuning dramatically and essentially saturating the general bound.

\vspace{0.3cm}
\noindent
{\large {\bf Acknowledgments}} We thank D.R.T. Jones for clarifications 
concerning ref.~\cite{EJ}. This work is supported in part by the Spanish
Ministry of Education and Science, through a M.E.C. project (FPA2001-1806).
The work of Irene Hidalgo has been supported by a FPU grant from the 
M.E.C.



\begin{thebibliography}{99}
%
\bibitem{veltman}
M.~J.~G.~Veltman,
Acta Phys.\ Polon.\ B {\bf 12} (1981) 437.
%
\bibitem{ewfits}
R.~Barbieri and A.~Strumia,
[hep-ph/0007265];
R.~Barbieri, A.~Pomarol, R.~Rattazzi and A.~Strumia,
[hep-ph/0405040].
%
\bibitem{KM}
C.~F.~Kolda and H.~Murayama,
JHEP {\bf 0007} (2000) 035 [hep-ph/0003170].
%
\bibitem{Calmet}
X.~P.~Calmet,
Eur.\ Phys.\ J.\ C {\bf 32} (2003) 121
[hep-ph/0302056].
%
\bibitem{EJ}
M.~B.~Einhorn and D.~R.~T.~Jones,
Phys.\ Rev.\ D {\bf 46} (1992) 5206.
%
\bibitem{LH}
N.~Arkani-Hamed, A.~G.~Cohen, E.~Katz and A.~E.~Nelson,
JHEP {\bf 0207} (2002) 034
[hep-ph/0206021].
%
\bibitem{BG}
R.~Barbieri and G.~F.~Giudice,
Nucl.\ Phys.\ B {\bf 306} (1988) 63.
%
\bibitem{top}
P.~Azzi {\it et al.}  [CDF Collaboration],
[hep-ex/0404010].
%
\bibitem{CS}
P.~Ciafaloni and A.~Strumia,
Nucl.\ Phys.\ B {\bf 494} (1997) 41
[hep-ph/9611204].
%
\bibitem{LEPh}
R.~Barate {\it et al.}  [LEP Collaborations],
Phys.\ Lett.\ B {\bf 565} (2003) 61
[hep-ex/0306033].
%
\bibitem{abuse}
C.~P.~Burgess and D.~London,
Phys.\ Rev.\ D {\bf 48} (1993) 4337
[hep-ph/9203216].
%
\bibitem{Alex}
T.~Gherghetta and A.~Pomarol,
Phys.\ Rev.\ D {\bf 67}, 085018 (2003)
[hep-ph/0302001].
%
\bibitem{CCIQ}
J.~A.~Casas, V.~Di Clemente, A.~Ibarra and M.~Quir\'os,
Phys.\ Rev.\ D {\bf 62} (2000) 053005
[hep-ph/9904295];
J.~A.~Casas and A.~Ibarra,
Nucl.\ Phys.\ B {\bf 618} (2001) 171
[hep-ph/0103065].
%
\bibitem{ftgalore}
B.~de Carlos and J.~A.~Casas,
Phys.\ Lett.\ B {\bf 309} (1993) 320
[hep-ph/9303291];
%
G.~W.~Anderson and D.~J.~Casta\~no,
Phys.\ Lett.\ B {\bf 347} (1995) 300
[hep-ph/9409419];
Phys.\ Rev.\ D {\bf 52} (1995) 1693
[hep-ph/9412322];
Phys.\ Rev.\ D {\bf 53} (1996) 2403
[hep-ph/9509212];
%
S.~Dimopoulos and G.~F.~Giudice,
Phys.\ Lett.\ B {\bf 357} (1995) 573
[hep-ph/9507282];
%
G.~W.~Anderson, D.~J.~Castano and A.~Riotto,
Phys.\ Rev.\ D {\bf 55} (1997) 2950
[hep-ph/9609463];
K.~Agashe and M.~Graesser,
Nucl.\ Phys.\ B {\bf 507} (1997) 3
[hep-ph/9704206];
%
P.~H.~Chankowski, J.~R.~Ellis and S.~Pokorski,
Phys.\ Lett.\ B {\bf 423} (1998) 327
[hep-ph/9712234];
%
R.~Barbieri and A.~Strumia,
Phys.\ Lett.\ B {\bf 433} (1998) 63
[hep-ph/9801353];
%
P.~H.~Chankowski, J.~R.~Ellis, M.~Olechowski and S.~Pokorski,
Nucl.\ Phys.\ B {\bf 544} (1999) 39
[hep-ph/9808275];
%
G.~L.~Kane and S.~F.~King,
Phys.\ Lett.\ B {\bf 451} (1999) 113
[hep-ph/9810374];
%
L.~Giusti, A.~Romanino and A.~Strumia,
Nucl.\ Phys.\ B {\bf 550} (1999) 3
[hep-ph/9811386];
%
J.~L.~Feng, K.~T.~Matchev and T.~Moroi,
Phys.\ Rev.\ Lett.\  {\bf 84} (2000) 2322
[hep-ph/9908309];
%
K.~Agashe,
Phys.\ Rev.\ D {\bf 61} (2000) 115006
[hep-ph/9910497];
%
M.~Bastero-Gil, G.~L.~Kane and S.~F.~King,
Phys.\ Lett.\ B {\bf 474} (2000) 103
[hep-ph/9910506];
%
A.~Romanino and A.~Strumia,
Phys.\ Lett.\ B {\bf 487} (2000) 165
[hep-ph/9912301];
%
G.~L.~Kane, J.~Lykken, B.~D.~Nelson and L.~T.~Wang,
Phys.\ Lett.\ B {\bf 551} (2003) 146
[hep-ph/0207168].
%
\bibitem{CEH}
J.~A.~Casas, J.~R.~Espinosa and I.~Hidalgo,
JHEP {\bf 0401}, 008 (2004)
[hep-ph/0310137].
%
\bibitem{Strumia}
A.~Strumia,
Phys.\ Lett.\ B {\bf 466} (1999) 107 [hep-ph/9906266];
R.~Barbieri, et al.
Nucl.\ Phys.\ B {\bf 663}, 141 (2003) [hep-ph/0208153];
D.~Marti and A.~Pomarol,
Phys.\ Rev.\ D {\bf 66}, 125005 (2002)
[hep-ph/0205034];
R.~Barbieri, G.~Marandella and M.~Papucci,
Phys.\ Rev.\ D {\bf 66}, 095003 (2002)
[hep-ph/0205280].
%
\bibitem{extG}
See {\em e.g.}
D.~Comelli and C.~Verzegnassi,
Phys.\ Lett.\ B {\bf 303} (1993) 277;
J.~R.~Espinosa and M.~Quir\'os,
Phys.\ Lett.\ B {\bf 302} (1993) 51
[hep-ph/9212305];
M.~Cveti\v c, D.~A.~Demir, J.~R.~Espinosa, L.~L.~Everett and P.~Langacker,
Phys.\ Rev.\ D {\bf 56} (1997) 2861
[Erratum-ibid.\ D {\bf 58} (1998) 119905]
[hep-ph/9703317];
P.~Batra, A.~Delgado, D.~E.~Kaplan and T.~M.~Tait,
[hep-ph/0309149].
%
\bibitem{extH}
M.~Drees,
Int.\ J.\ Mod.\ Phys.\ A {\bf 4} (1989) 3635;
J.~R.~Ellis, J.~F.~Gunion, H.~E.~Haber, L.~Roszkowski and F.~Zwirner,
Phys.\ Rev.\ D {\bf 39} (1989) 844;
P.~Binetruy and C.~A.~Savoy,
Phys.\ Lett.\ B {\bf 277} (1992) 453.
J.~R.~Espinosa and M.~Quir\'os,
Phys.\ Lett.\ B {\bf 279} (1992) 92;
Phys.\ Rev.\ Lett.\  {\bf 81} (1998) 516
[hep-ph/9804235];
G.~L.~Kane, C.~F.~Kolda and J.~D.~Wells,
Phys.\ Rev.\ Lett.\  {\bf 70} (1993) 2686
[hep-ph/9210242].
%
\bibitem{NMSSM}
M.~Bastero-Gil, C.~Hugonie, S.~F.~King, D.~P.~Roy and S.~Vempati,
Phys.\ Lett.\ B {\bf 489} (2000) 359
[hep-ph/0006198].
%
\bibitem{hard}
K.~Harada and N.~Sakai,
Prog.\ Theor.\ Phys.\  {\bf 67} (1982) 1877;
D.~R.~Jones, L.~Mezincescu and Y.~P.~Yao,
Phys.\ Lett.\ B {\bf 148} (1984) 317;
I.~Jack and D.~R.~Jones,
Phys.\ Lett.\ B {\bf 457} (1999) 101
[hep-ph/9903365];
L.~J.~Hall and L.~Randall,
Phys.\ Rev.\ Lett.\  {\bf 65} (1990) 2939;
F.~Borzumati, G.~R.~Farrar, N.~Polonsky and S.~Thomas,
Nucl.\ Phys.\ B {\bf 555} (1999) 53
[hep-ph/9902443];
S.~P.~Martin,
the
Phys.\ Rev.\ D {\bf 61} (2000) 035004
[hep-ph/9907550].
%
\bibitem{Brignole}
A.~Brignole, F.~Feruglio and F.~Zwirner,
Nucl.\ Phys.\ B {\bf 501} (1997) 332
[hep-ph/9703286].
\bibitem{Polonsky}
N.~Polonsky and S.~Su,
Phys.\ Lett.\ B {\bf 508} (2001) 103
[hep-ph/0010113];
Phys.\ Rev.\ D {\bf 63} (2001) 035007
[hep-ph/0006174].
%
\bibitem{BCEN}
A.~Brignole, J.~A.~Casas, J.~R.~Espinosa and I.~Navarro,
Nucl.\ Phys.\ B {\bf 666} (2003) 105 [hep-ph/0301121].
%
\bibitem{splitsusy}
N.~Arkani-Hamed and S.~Dimopoulos,
[hep-th/0405159];
G.~F.~Giudice and A.~Romanino,
[hep-ph/0406088];
N.~Arkani-Hamed, S.~Dimopoulos, G.~F.~Giudice and A.~Romanino,
[hep-ph/0409232].
%
\bibitem{GM}
G.~F.~Giudice and A.~Masiero,
Phys.\ Lett.\ B {\bf 206} (1988) 480.
%
\end{thebibliography}
\end{document}